# Comparison of experimental and theoretical results to define centrality of heavy ion collisions

*Z. Wazir*[1], M. K. Suleymanov[1,2], E. U. Khan[1], Mahnaz Q. Haseeb[1], M. Ajaz[1], K. H. Khan[1]*
[1]CIIT, Islamabad (Pakistan), [2]JINR, Dubna (Russia)
*Corresponding author: zafar_wazir@comsats.edu.pk

## Abstract:

Using the simulation data coming from the cascade model, we have studied the behavior of event number as a function of impact parameter-b and a number of all charged particles- $N_{ch}$ for light and heavy nuclei at different energies. We have seen that for light nuclei, a number of all charged particles-$N_{ch}$ could be used to fix the centrality. But for heavy nuclei we have got strong initial energy and mass dependences and the results for impact parameter factor dependences and ones for a number of all charged particles differ. So for heavy nuclei, a number of charged particles-$N_{ch}$ could not be use to fix the centrality.

## 1. Introduction.

To fix the baryon density of nuclear matter the centrality experiments are usually used. It is considered as best tool to reach the Quark Gluon Phase (QGP) [1] of nuclear matter under extreme conditions. Studying the different characteristics of events as a function of the centrality [2-4] in JINR (Dubna), CERN (Geneva), BNL (New-York), and SIS (Darmstadt) could give new information about the properties of nuclear matter which could appear under extreme conditions. On the other hand the centrality of collisions cannot be defined directly in the experiment. In different experiments the values of the centrality are defined [5-9]as a number of identified protons , projectiles' and targets' fragments , slow particles , all particles , as the energy flow of the particles with emission angles $\theta \cong 0^{\circ}$ or with $\theta \cong 90^{\circ}$ . Apparently, it is not simple to compare quantitatively the results on centrality-dependences obtained in literature while on the other hand the definition of centrality could significantly influence the final results. May be this is a reason, why we could not get a clear signal on new phases of strongly interacting matter, though a lot of interesting information has been given in those experiments.

During last several years some results of the central experiments are discussed which demonstrate the point of regime change and saturation on the behavior of some characteristics of the events as a function of the centrality [10]. It is supposed that these phenomena could be connected with fundamental properties of the strongly interacting mater and could reflect the changes of its states (phases).

Let us take some examples. In paper [11] the results from BNL experiment E910 on pion production and stopping in proton-Be, Cu, and Au collisions as a function of centrality at a beam momentum of 18 GeV/c are presented. The centrality of the collisions is characterized using the measured number of «grey» tracks, $N_{grey}$, and a derived quantity$\nu$, the number of inelastic nucleon-nucleon scatterings suffered by the projectile during the collision. In Fig. 1 is plotted the values of average multiplicity for $\pi^-$ - mesons (<$\pi^-$ Multiplicity>) as a function of Ngrey and $\nu$ for the three different targets. One can observe that <$\pi^-$ Multiplicity> increases approximately proportionally to $N_{grey}$ and $\nu$ for all three targets at small values of $N_{grey}$ or $\nu$ and saturates with increasing $N_{grey}$ and $\nu$ in the region of more high values of $N_{grey}$ and$\nu$.

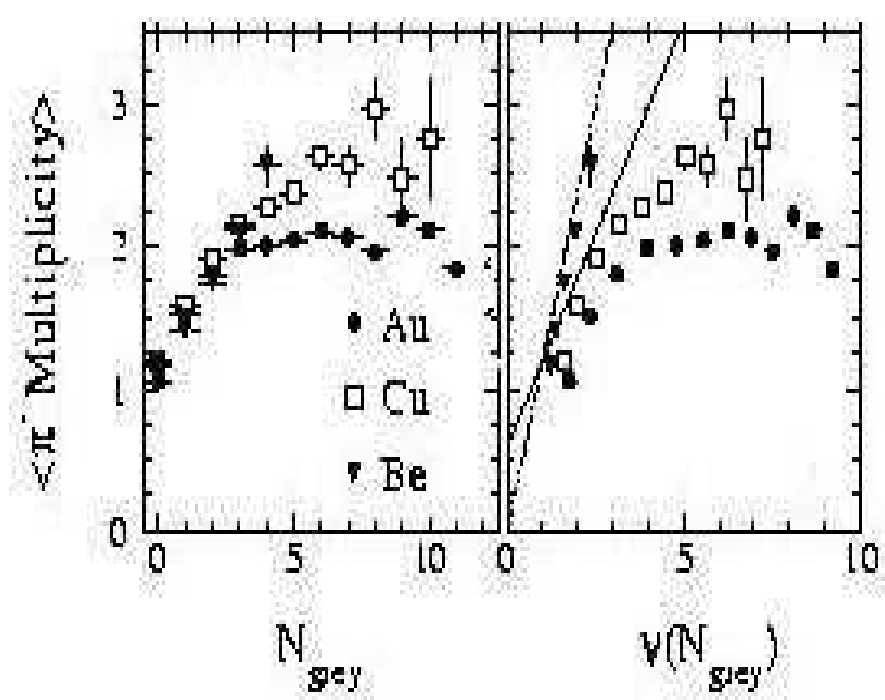

Fig.1**.** The average multiplicity of the $\pi^-$ -mesons produced in proton-Be, Cu, and Au collisions as a function of centrality at a beam momentum of 18 GeV/c. Solid line demonstrates the results coming from the WN-model.

BNL E910 has measured Λ production as a function of collision centrality for 17.5 GeV/c p–Au reactions [12]. They observed that the measured Λ yield increases with centrality faster than saturates. This Collaboration has obtained the same results for $K^0_s$ and $K^+$ -mesons emitted in p+Au reaction.

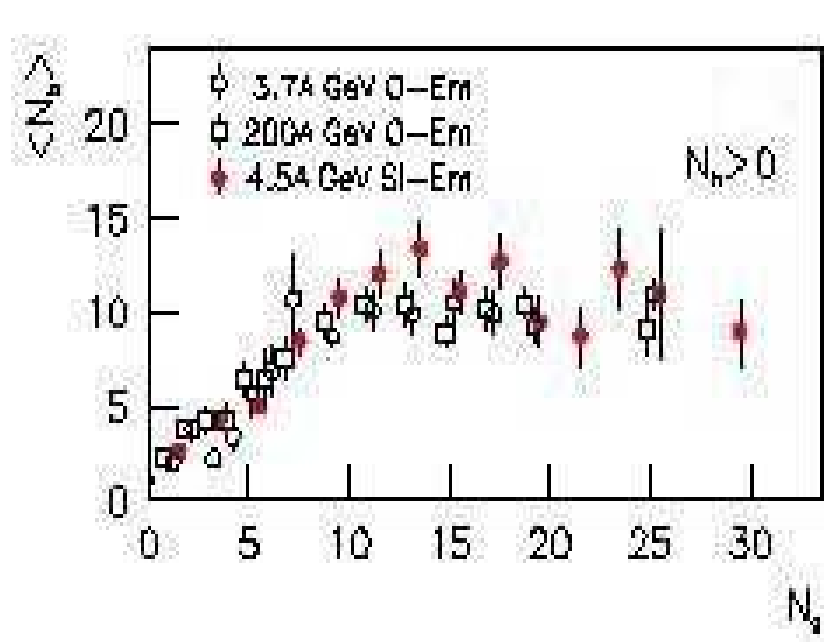

Fig. 2: $N_g$ –dependences of $< N_b >$ for different reactions.

Fig. 2 is a plot of multiplicity of grey particles - Ng–dependences verses $< N_b >$ average multiplicity of b-particles for different reactions taken from [13]. One can see that the values of $< N_b >$ increase with $N_g$ in the region of the values of Ng< 8. Than the values of the $< N_b >$ saturate in the region $N_g \geq 8$ as well as in Ref. [14] .

# 2. Main Results.

The main results of central experiments are: The regime change has been observed: at some values of centrality (as critical phenomena); for hadron-nucleus, nucleus-nucleus and even in ultra relativistic heavy ion collisions; in the energy ranges from SIS energy up to RHIC; almost for all particles; after the point of regime change, saturation is observed; the existing simple models cannot explain the effect. If the regime change takes place unambiguously two times, this would surely be the most direct experimental evidence seen to observe the QCD critical point and phase transition. But the central experiments could not confirm it. One of the reasons of it may be is the not correct definition of the centrality**.** So it is very actually study the connections between the different methods for fixing the centrality and looking for the new possibilities to fix the centrality especially in heavy ion collisions where the formation of QGP is expected. The main goal of our paper is to study the connections between different methods offered to fix the centrality and search for new methods to fix the centrality.

# 3. Method.

To reach our goal we use the simulation data coming from the Cascade Model (CM). We therefore started from CM that usually for a chosen variable to fix centrality it is supposed that its values have to increase linearly with a number of colliding nucleons or baryon density of the nuclear matter. The simplest mechanism that could give this dependence is the cascade approach. So, we have used one of the versions of the cascade model – Dubna Intra-Nuclear Cascade Model (DINCM) to simulate events at different energies and mass colliding hadrons and nuclei. The code is written by F.G. Geregy and J.J. Musulmanbekov [15] and was modified by S.Yu. Shmakovand V.V. Uzhinskii in 1993. The DINCM is used to calculation of nucleus-nucleus inelastic interactions at energies up to 20 A GeV.

The following reactions were considered by us: He+He; C+C; Au+Au at the energies; 1; 6; 12; 18 A GeV/c.

We considered two variables two fix the centrality: -- impact factor *b,* which could not define experimentally; -- a number of all charged particles $N_{ch}$, which could be defined experimentally.

# 4. Results.

The behaviour of the normalized event number d*N/db* as a function of *b* (fig. 3a ) and the $dN/dN_{ch}$ as a function of *b* (fig.3b) for *He+He* reactions at different initial energies are shown in this pictures. One can see (fig. 3a) that the behaviour of the distributions don't depend on the energy of the colliding nuclei for most central (*b=0*), central and semi central collisions ($0 < b < 3$). We can see some mass dependence for the peripheral collisions ($b>3$). We can also say that the there are 2 regions on the behaviour of the *dN/db* as a function of the *b*. In first region – $b < 3$ the values of d*N/db* great than in region with $b > 3$.

The behaviour of the event number as a function of $N_{ch}$ *has the* stronger energy dependences. We can say that at energies equal and great than 6 GeV we can find some analogies between the behaviour of the distribution of the events as a function of the *b* and $N_{ch}$. It means in these cases the last could be use to fix the centrality instate the *b*.

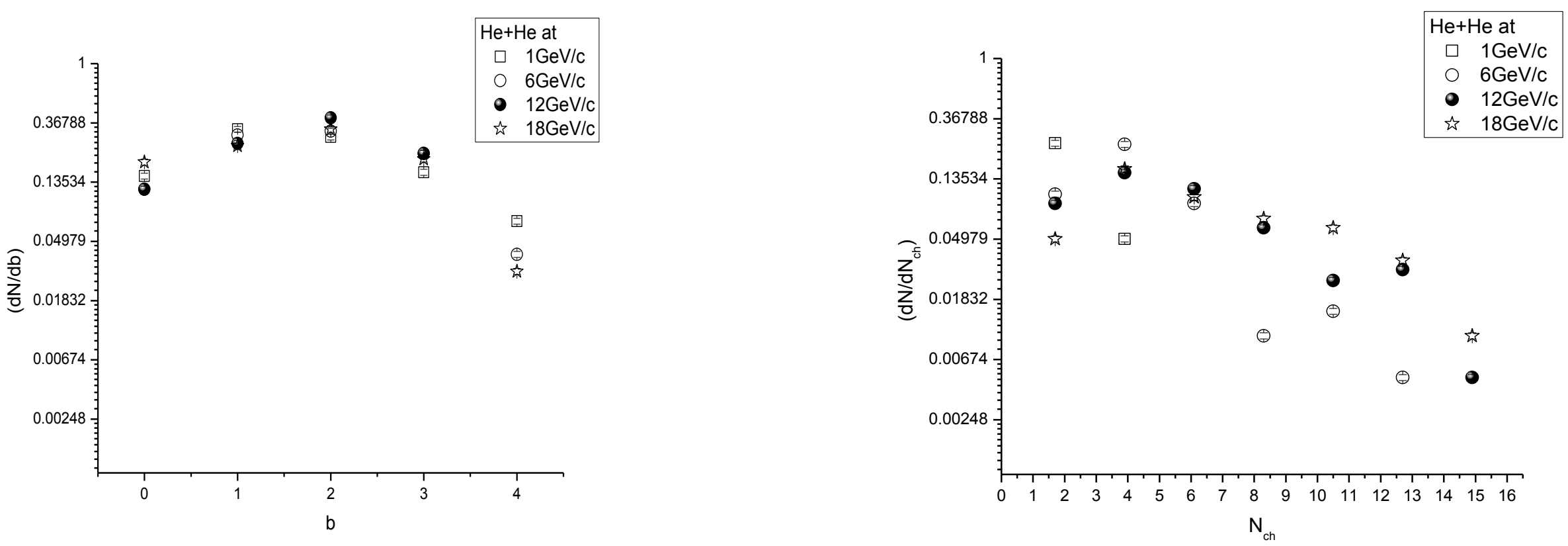

Fig.3 a-b. The *b*- (a) and $N_{ch}$ - (b) dependences of normalized event numbers for *He+He* interactions coming from DINCM

Fig. 4 is shown the behavior of the normalized event number *dN/db* as a function of *b* (fig. 4a) and $N_{ch}$ (fig. 4b) as a function of *b* for *CC*–reactions at the chosen values of the energies. There is some energy dependence for the behavior of *dN/db* as a function of the *b* in the region of momentum great that 12 AGeV/c. For these reactions there 3 region on the behavior of the *N* as a function of the *b*: *b=0* most central collisions; $0< b <5$ central and semi central collisions and $b>5$ peripheral collisions. So one can say that with increasing the mass of the colliding particles DINCM give some energy dependence for the behavior event number as a function of *b.*

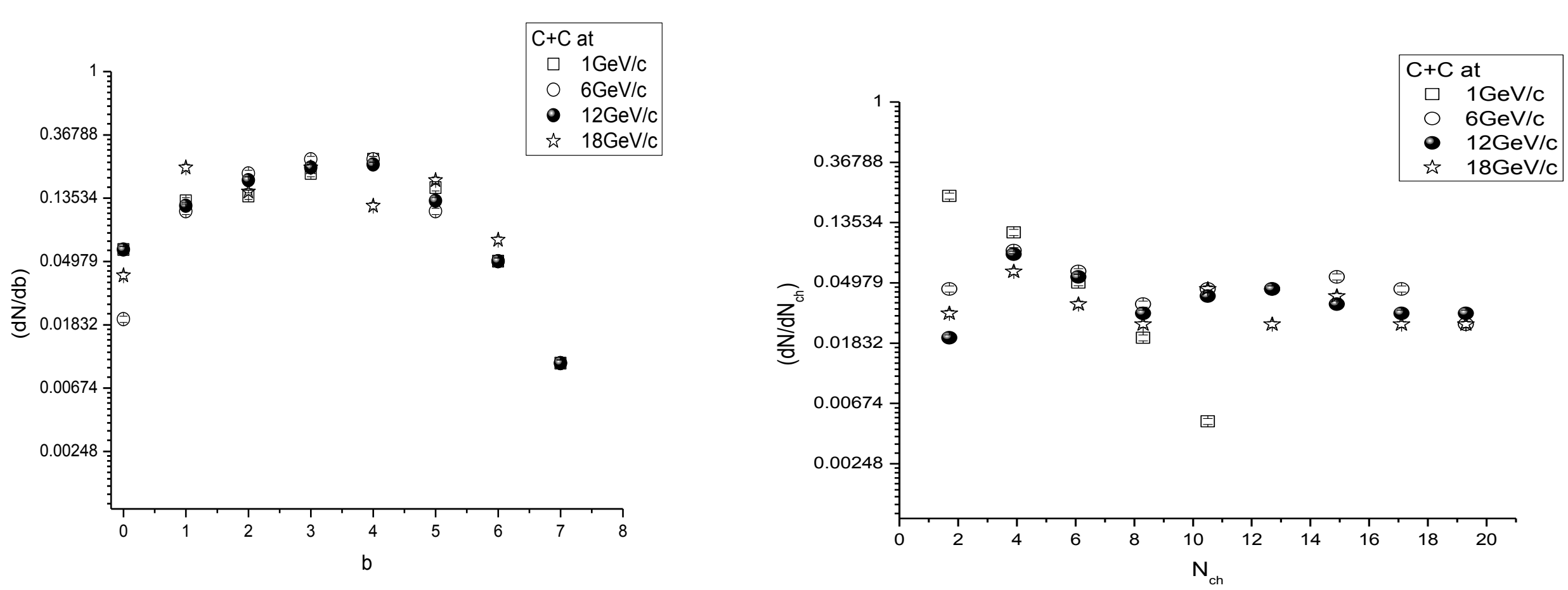

Fig.4 a-b. The *b*- (a) and $N_{ch}$ - (b) dependences of normalized event numbers for *C+C*- interactions coming from DINCM

Again we can say that the behavior of the event number as a function of $N_{ch}$ has the stronger energy dependences. The fluctuation in the behavior of the event number as a function of $N_{ch}$ increase and it is very difficult to find some analogies between the behavior of the distribution of the events as a function of the b and $N_{ch}$. So it means that it will be very difficult to use the last to fix the centrality instead the *b*. The same result we can get for the heavy ion collisions.

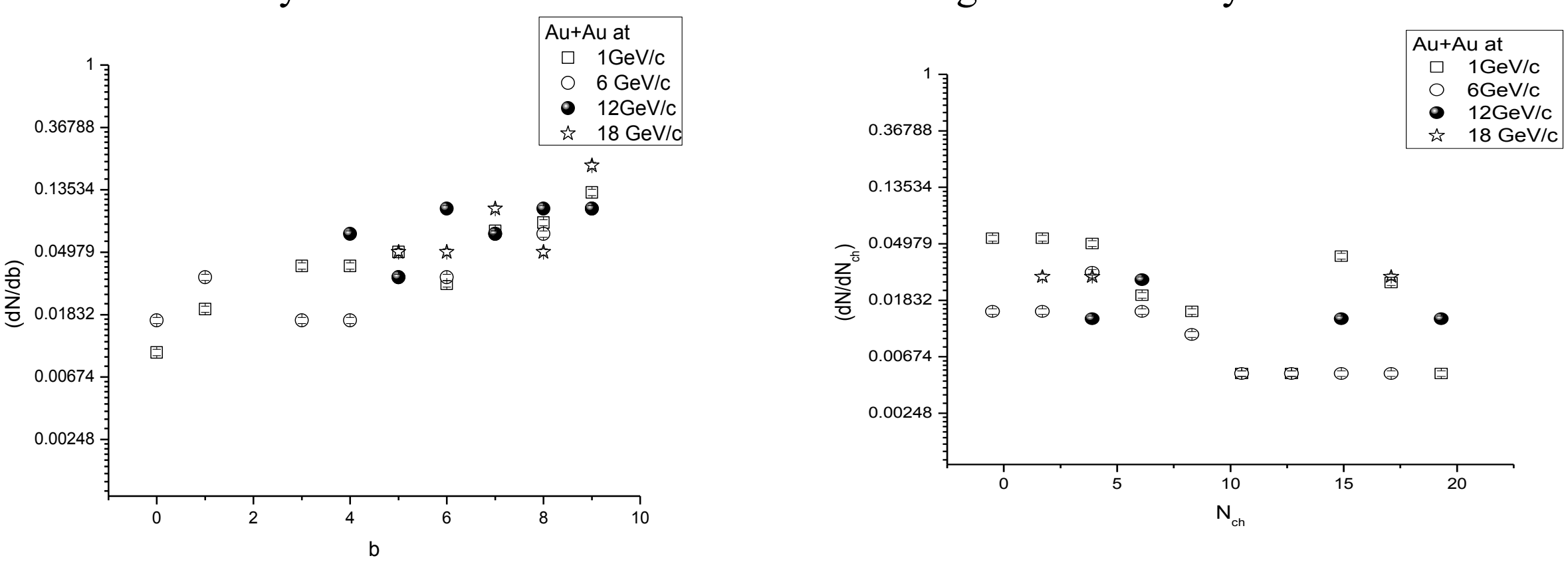

Fig.5 a-b. The *b*- (a) and $N_{ch}$ - (b) dependences of normalized event numbers for *Au+Au*- interactions coming from DINCM

For *Au+Au* reactions at different energies the *b and Nch* dependences of normalized event number *dN/db* and $dNdN_{ch}$ are shown in the Fig.5a and Fig.5b. We can see the strong dependence for the behavior of event numbers as a function of the centrality. This picture also indicates different regions for the behavior of event numbers as a function of the *b* (fig. 5a) but different depend on the energy. We cannot find any analogy for the behavior of the distributions in figures 5a and 5b. It can mean that for heavy nuclear interactions $N_{ch}$ *is* not good variable to fix the centrality.

## Conclusion.

The behaviour of the normalized event number as a function of impact parameter and a number of all charged particles for *He+He-, C+C- and Au+Au*- reactions at different initial energies coming from cascade model are point that the for light nuclei a number of all charged particles could be used to fix the centrality. For heavy nuclei we have got strong initial energy and mass dependences and the results for impact factor dependences and ones for a number of all charged particles differ. So in this cases a number of events could not be use to fix the centrality.